\def\draftversion{false}
\RequirePackage{ifthen}
\ifthenelse{\equal{\draftversion}{true}}{
  \documentclass[aps,prl,10pt,galley,amsmath,amssymb, 
                 superscriptaddress]{revtex4}
}{
  \documentclass[aps,prl,10pt,twocolumn,amsmath,amssymb, 
                 superscriptaddress,longbibliography]{revtex4-1}
}

\ifthenelse{\equal{\draftversion}{true}}{
  \marginparwidth 2.7in
  \marginparsep 0.5in
  \newcounter{comm} 
  \def\commnext{\stepcounter{comm}}
  \def\commtext{{\bf\color{blue}[\arabic{comm}]}}
  \def\commmar{{\bf\color{blue}[\arabic{comm}]}}
  \def\dvm#1{\commnext\marginpar{\small DV\commmar: #1}\commtext}
  \def\khm#1{\commnext\marginpar{\small KH\commmar: #1}\commtext}
  \def\hsm#1{\commnext\marginpar{\small HS\commmar: #1}\commtext}
  \def\mlab#1{\marginpar{\small\bf #1}}
  
}{
  \def\dvm#1{}
  \def\khm#1{}
  \def\hsm#1{}
  \def\mlab#1{}
  
}

\usepackage{soul}  

\usepackage[]{graphicx}
\usepackage[]{verbatim}
\usepackage[dvipsnames]{xcolor}
\usepackage{tabularx}

\usepackage{amsfonts}
\usepackage{amsmath}
\usepackage{amssymb}
\usepackage{bm}
\usepackage{graphicx}
\usepackage[breaklinks,colorlinks=true,citecolor=blue]{hyperref}
\usepackage{mathrsfs}
\usepackage[lofdepth,lotdepth,caption=false]{subfig}
\usepackage{varwidth}
\usepackage{wrapfig}
\usepackage{times}
\usepackage{longtable}
\usepackage{multirow}
\usepackage{tikz}
\usepackage{enumerate}

\makeatletter

\AtBeginDocument{\@ifpackageloaded{natbib}{\ifNAT@numbers\if@filesw\immediate\write\@auxout{\string\global\string\NAT@numberstrue}\fi\fi}{}}
\makeatother
\begin{document}

\title{
  Molecular Mott state in the deficient spinel GaV$_4$S$_8$
}

\author{Heung-Sik Kim}
\affiliation{Department of Physics and Astronomy, Rutgers University, Piscataway, New Jersey 08854-8019, USA}
\affiliation{Department of Physics, Kangwon National University, Chuncheon 24341, Korea}

\author{Kristjan Haule}
\affiliation{Department of Physics and Astronomy, Rutgers University, Piscataway, New Jersey 08854-8019, USA}

\author{David Vanderbilt}
\affiliation{Department of Physics and Astronomy, Rutgers University, Piscataway, New Jersey 08854-8019, USA}

\begin{abstract}

In this study, we investigated theoretically the Mott-insulating phase of a deficient spinel chalcogenide GaV$_4$S$_8$, which is known to form a tetrahedral V$_4$S$_4$ cluster unit that results in molecular orbitals (MOs) with a narrow bandwidth in the noninteracting limit. 
We used a cluster extension of charge self-consistent embedded dynamical mean-field theory to study the impact of strong intra-cluster correlations on the spectral properties as well as the structural degrees of freedom of the system. We found that the strong tetrahedral clustering renders the atomic Mott picture ineffective, and that the resulting MO picture is essential to describe the Mott phase. It was also found that, while the spectral properties can be qualitatively described by the truncation of the Hilbert space down to the lowest-energy MO, a proper description of the structural degrees of freedom requires the inclusion of multi-MO correlations that span a larger energy window. Specifically, we found that the lowest-energy MO description overemphasizes the clustering tendency, while the inclusion of the Hund's coupling between the lower- and higher-energy MOs corrects this tendency, bringing the theoretically predicted crystal structure into good agreement with the experiment. 

\end{abstract}

\maketitle

Intermetallic covalency in transition-metal chalcogenides or oxides often leads to the formation of density waves or transition-metal clustering~\cite{Whangbo1992}. While this typically results in a reduction of Fermi surface, more dramatic changes may happen in correlated systems such as VO$_2$~\cite{Morin1959,Qazilbash2007,Biermann2004,Brito2016} or 1$T$-phases of TaS$_2$ and NbSe$_2$~\cite{Wilson1975,DISALVO1977,Fazekas1979,Nakata2016,Calandra2018}. Another interesting class of materials is ternary deficient spinel chalcogenides $AM_4X_8$ ($A$ = Al, Ga, Ge; $M$ = Ti, V, Nb, Mo, Ta; $X$ = S, Se), where the four $M$ sites form a tetrahedral cluster and drive the system to be Mott insulating~\cite{YAICH19849,Pocha2000,Pocha2005,Johrendt1999,Helen2006CM,CHUDO2006,Vaju2008,Dorolti2010,HSK2014NC}. Among this family, GaV$_4$S$_8$ has been actively studied recently because of the existence of a rhombohedral polar ({\it i.e.}, with nonzero bulk electric polarization) phase with significant magnetoelectric coupling and the formation of a skyrmion crystal below $T_C$ = 13 K~\cite{Kezsmarki2015,Ruffe1500916,Widmann2017}. 
%
%
Based on this observation and the strong V$_4$ clustering, This system has been suggested to be a Mott insulator with the V$_4$ molecular orbitals (MO) comprising the correlated subspace. It seems likely that, as in the example of VO$_2$, the electron-lattice coupling in GaV$_4$S$_8$ can be modified by electron correlations in a non-trivial manner, which may affect the nature of the low-temperature multiferroic phase~\cite{Kezsmarki2015,Ruffe1500916,Widmann2017}. 

Dynamical mean-field theory (DMFT) has become a standard tool for tackling such correlated materials in an {\it ab-initio} manner~\cite{DMFTreview1,DMFTreview2,DMFTreview3}. The cluster extension of the conventional single-site DMFT~\cite{DMFTreview2,KotliarCDMFT2001} can be used to systematically increase the range of spatial correlations, extending the notion of locality from an atomic site to a cluster. 
However, the exponential scaling of the computational cost becomes an issue at this point; the number of cubic $t_{\rm 2g}$-orbitals in the V$_4$ cluster is 12, and directly tackling such problem is extremely challenging even with the use of the most state-of-the-art impurity solvers such as continuous-time Monte Carlo~\cite{HauleCTQMC,PatrickCTQMC,CTQMCreview}. Because of this difficulty, a proper {\it ab-initio} study of the Mott phase of GaV$_4$S$_8$, fully incorporating lattice and charge degrees of freedom, has not yet appeared.

\begin{figure}
  \centering
  \includegraphics[width=0.45\textwidth]{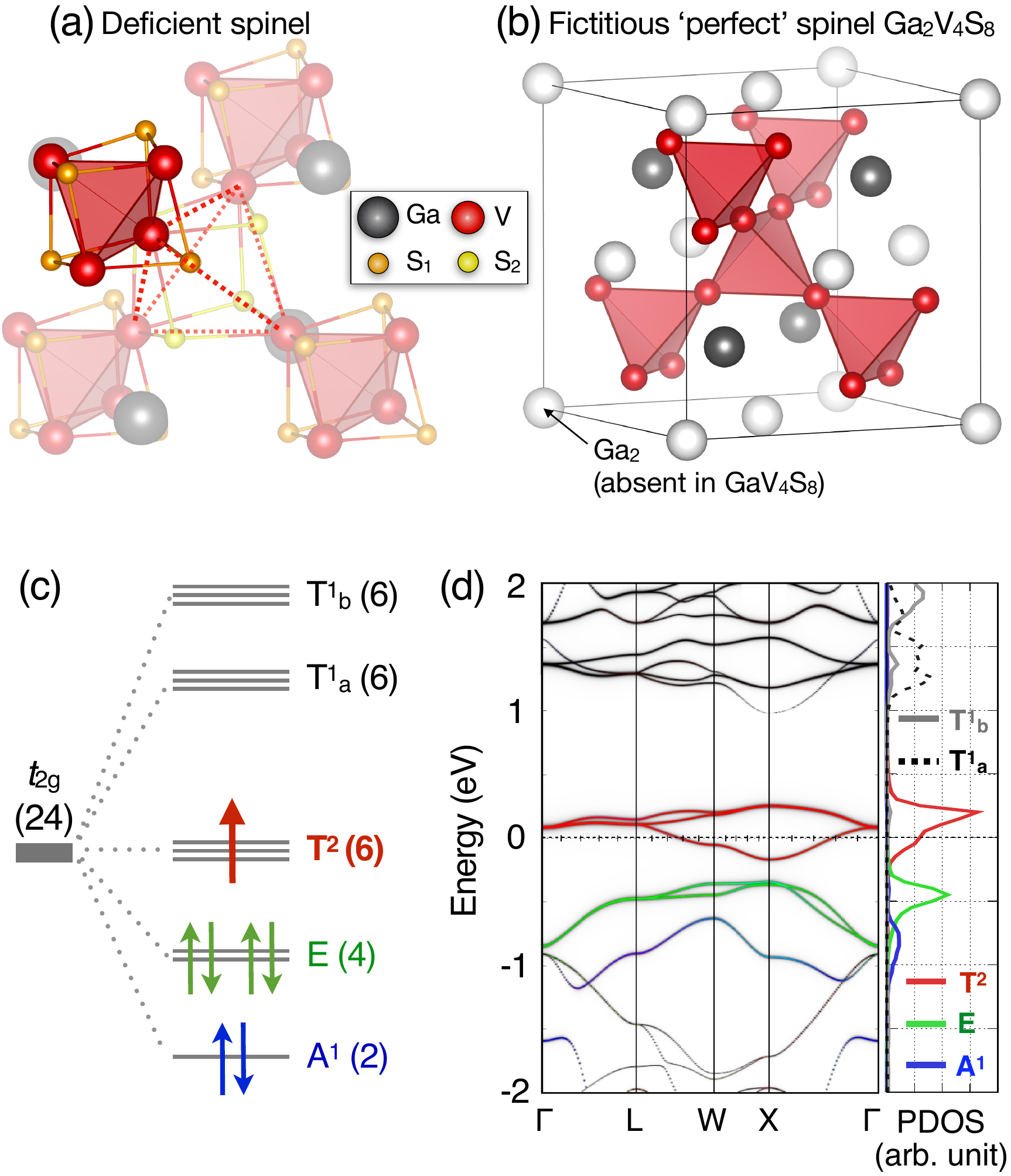}
  \caption{
  (a) Crystal structure of the deficient spinel GaV$_4$S$_8$ in the cubic phase, in comparison 
  with a fictitious perfect spinel Ga$_2$V$_4$S$_8$ illustrated in (b). Note the inter-cluster V-V bonds 
  depicted in red dashed lines in (a), and white Ga sites in (b) which are absent in deficient spinel structure (a). 
  (c) Splitting of 12 atomic $t_{\rm 2g}$ orbitals at 4 V sites in the V$_4$S$_4$ cluster into 
  the molecular-orbital (MO) states. Seven electrons in the (V$_4$)$^{13+}$ cluster occupy the 
  singlet $A^1$, doublet $E$, and triplet $T^2$ states, as shown the diagram. 
  (d) MO-projected fat-band representation and density of state (PDOS) plots of GaV$_4$S$_8$ from the DFT
  results (without $U$). 
  }
  \label{fig:dft}
\end{figure}

Hence, in this study, we have studied the Mott phase of GaV$_4$S$_8$ in the high-temperature cubic (non-polar) phase above $T$ = 45~K, specifically focusing on the occurrence of the Mott phase via the MO formation and its impact on the structural degrees of freedom. 
We employed fully charge self-consistent DMFT with cluster MO bases applied to the tetrahedral cluster of four V sites, starting from the simplest model containing only the lowest-energy MO ($T^2$ in Fig.~\ref{fig:dft}) and progressively enlarging the correlated Hilbert space to include the majority of $t_{\rm 2g}$ states in the $V_4$ cluster ($T^2+E+T^1_a$). 
Our cluster MO-DMFT prediction was compared to most standard as well as advanced density functional theory (DFT) exchange-correlation functionals, including SCAN meta-GGA~\cite{SCAN} and HSE hybrid functionals~\cite{HSE06,HSE06-2}. While these all fail to predict an insulating phase, our cluster calculation opens a gap very naturally, thus demonstrating that the MO picture is essential for describing the Mott phase. Surprisingly, the V$_4$S$_4$ clustering is strongly affected by the strength of the Hund's coupling at the V sites. The DMFT approach applied to this compound yields qualitatively different results compared to those obtained from DFT or DFT+$U$~\cite{Sieberer2007}, demonstrating its power in tackling correlated systems with multisite clusters. 

{\color{NavyBlue}\it Computational tools.} 
To incorporate the electronic and structural degrees of freedom on an equal footing, we employed a state-of-the-art DFT+embedded DMFT code~\cite{Dmft,eDMFT} which allows relaxation of internal atomic coordinates. In DMFT the experimental lattice parameter reported in Ref.~\onlinecite{Pocha2000} was employed, and optimizations of internal atomic coordinates were done using DMFT forces~\cite{Haule2015FE,Force2016}. The hybridization-expansion continuous-time quantum Monte Carlo method~\cite{HauleCTQMC,PatrickCTQMC} was employed as the impurity solver. The atomic on-site Coulomb interactions were unitarily transformed and projected onto the MO basis, where the impurity hybridization function has a more appropriate form for the impurity solver~\footnote{Details of this transformation and its implementation in the DFT+embedded DMFT code are discussed in the Supplementary Material (SM). Therein it is argued that intracluster Coulomb repulsions in this system should be insignificant and can be ignored~\cite{Haule_exactDC}. Note that the DFT+embedded DMFT code runs based on {\sc wien2k} package~\cite{wien2k}. Choices of $U$- and $J$-values in this DMFT implementation was discussed in Ref.~\onlinecite{Birol2014}. The Vienna {\it ab-initio} Simulation Package ({\sc vasp})~\cite{VASP1,VASP2} was used for independent structural optimizations at the DFT level.}.

\begin{figure}
  \centering
  \includegraphics[width=0.475\textwidth]{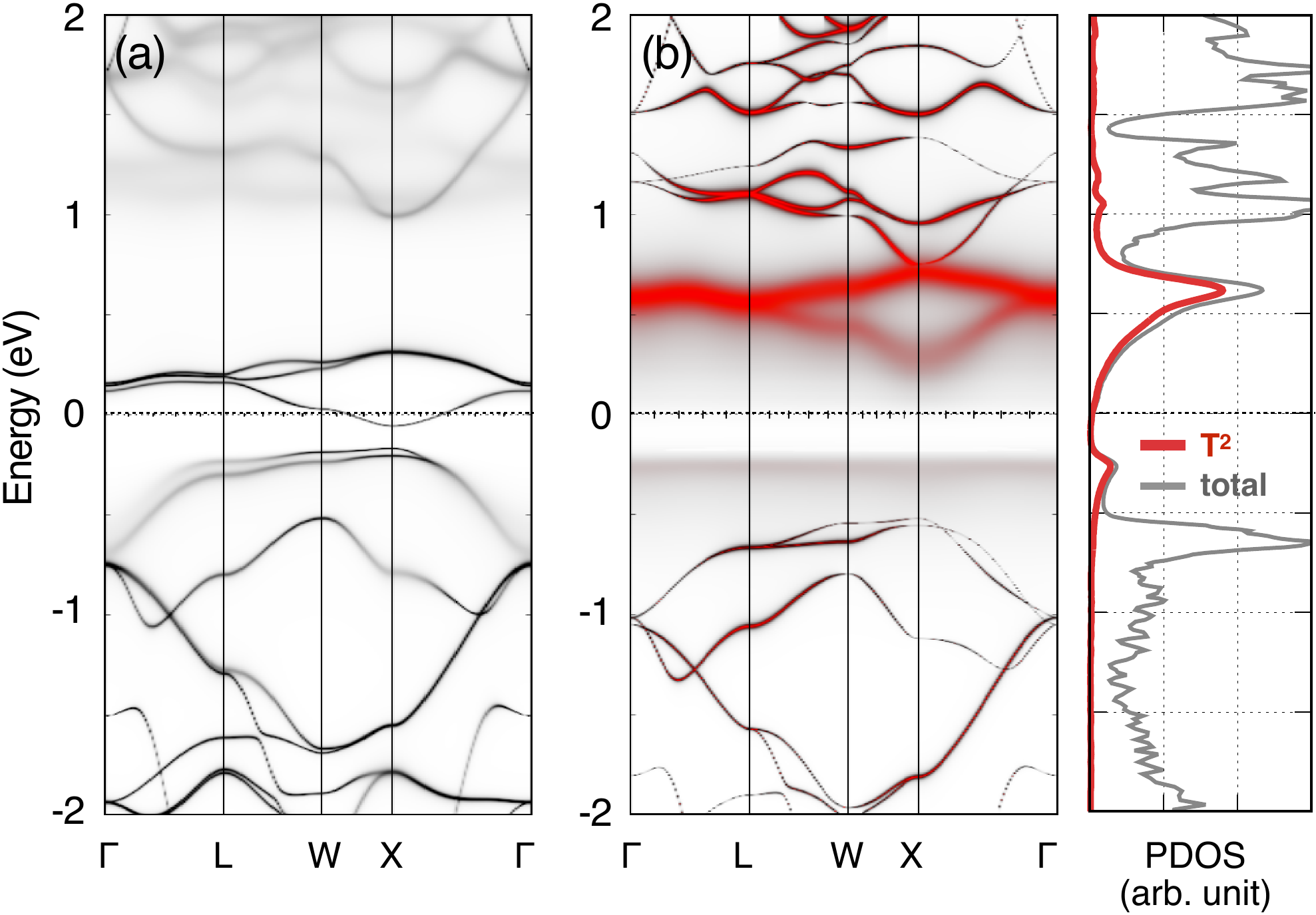}
  \caption{
  (a) A plot of the single-site DMFT spectral function with atomic V $t_{\rm 2g}$ states chosen as the
  correlated subspace ($U_d$ = 6 eV, $J_{\rm H}$ = 0.8 eV, T = 232K), showing a robust metallic character. 
  (b) MO-DMFT spectral function and PDOS with the MO-$T^2$ states as the correlated subspace 
  ($U_d$ = 6 eV, T = 232K). The red hue in the spectral function plot depicts the 
  character of the MO-$T^2$ states.   
  }
  \label{fig:MOT1}
\end{figure}

{\color{NavyBlue}\it Crystal structure and MO formation.} Fig.~\ref{fig:dft}(a) shows the crystal structure of cubic GaV$_4$S$_8$. Compared to the fictitious non-deficient spinel Ga$_2$V$_4$S$_8$ shown in Fig.~\ref{fig:dft}(b), half of the Ga sites (white Ga$_2$ sites in the figure) are missing in GaV$_4$S$_8$, which breaks the inversion symmetry (space group $F\bar{4}3m$) and allows the clustering of V and half of S (S$_1$ sites in the figure). This gives rise to MOs formed out of the 12 atomic $t_{\rm 2g}$ orbitals in the V$_4$ cluster, as depicted in Fig.~\ref{fig:dft}(c), where the 12 orbitals are split into 5 irreducible representations of the cubic $T_d$ point group, specifically $A^1 \oplus E \oplus T^{2} \oplus 2T^{1}$ (two $2T^{1}$ denoted as $T^{1}_{a,b}$ in the diagram). Note that the charge configuration is (V$_4$)$^{13+}$, so there are 7 electrons left in the cluster, fully occupying the singlet $A^1$ and doublet $E$ and filling one electron in the $T^2$ triplet, as shown in Fig.~\ref{fig:dft}(c). The result of a DFT calculation (without including $U$) is shown in Fig. \ref{fig:dft}(d), showing MO-projected fat bands and partial density of states (PDOS) where blue, green, and red colors depict the MO-$A^1$, $E$, and $T^{2}$ orbital characters respectively. The MOs can be seen to be well separated in energy and show a narrow bandwidth because of the strong clustering; compared to the size of MO splitting, which is on the order of $\sim$ 1 eV, the magnitude of the intercluster electron hopping is at most $\sim$ 0.02 eV~\cite{Camjayi2012}. This implies that the MO orbitals can be a reasonable basis set for the following MO-DMFT calculations.

\begin{figure*}
  \centering
  \includegraphics[width=0.98\textwidth]{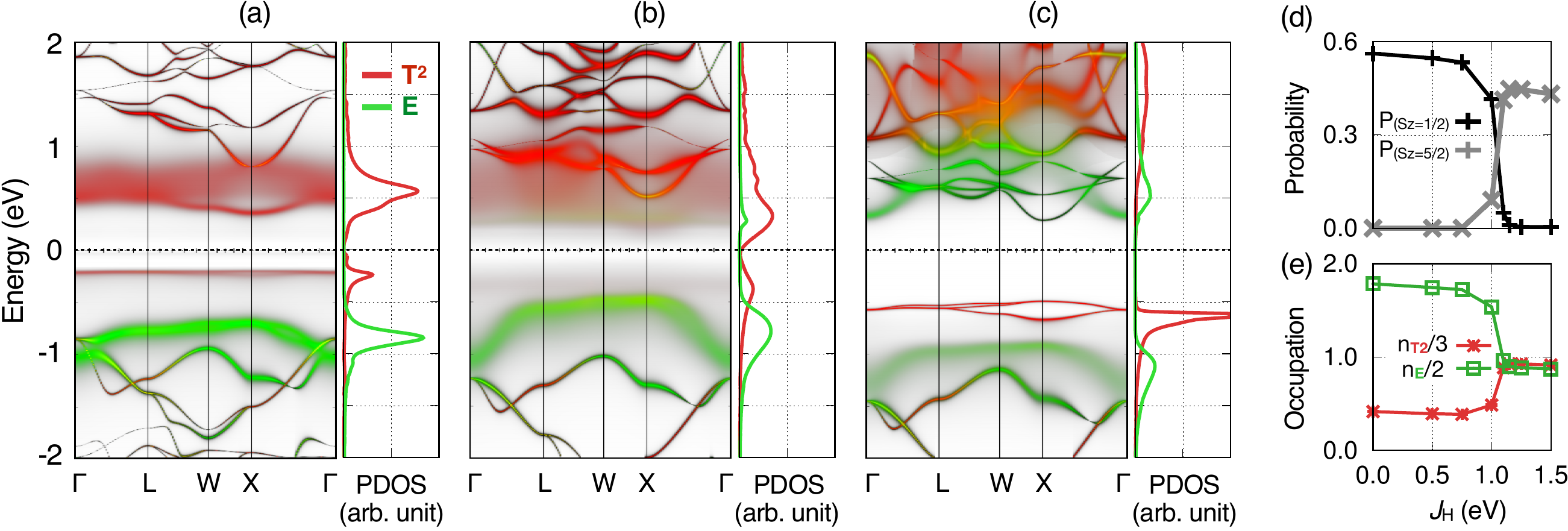}
  \caption{
  (a-c) Spectral functions and PDOSs with MO-$T^2$ and $E$ states as correlated subspaces, where the red and green hues
  depicting the $T^2$ and $E$ characters respectively. The Hund's coupling $J_{\rm H}$ for the correlated $T^2 \oplus E$ subspace 
  is varied from 0 to (a) 0.5 eV, (b) 1.0 eV, and (c) 1.5 eV. Evolution of (d) the probabilities of $S$ = 1/2 and 5/2 configurations and 
  (e) electron occupations in the $T^2$ (red) and $E$ (green) states as a function of $J_{\rm H}$. 
  }
  \label{fig:MOT2}
\end{figure*}

{\color{NavyBlue}\it Single-site vs.\ cluster MO DMFT.} 
Fig.~\ref{fig:MOT1} shows the comparison between the results from the conventional single-site DMFT and the simplest $T^2$-MO-DMFT calculations ($T$ = 232 K)~\footnote{See SM for the details of the single-site DMFT.}. In the latter scheme, one treats the partially-filled $T^2$ triplet MO as the correlated subspace. Note that choosing the $T^2$ only as the correlated subspace is the simplest cluster-type approximation, but it already yields a completely different result compared to the single-site DMFT. Fig.~\ref{fig:MOT1}(a) shows the $k$-dependent spectral function from the single-site DMFT calculation, employing the atomic V $t_{\rm 2g}$-orbitals as the correlated subspace with an on-site Coulomb repulsion of $U$ = 6 eV, appropriate for the V $t_{\rm 2g}$ set of quasi-atomic orbitals. A metallic band structure is clearly visible around the Fermi level, similar to the DFT result (Fig.~\ref{fig:dft}(d)), due to the strong hybridization between the intra-cluster V sites and the mixed valence occupancy ($d^{1.75}$ per V). Increasing the $U$ value within the single-site DMFT did not induce a qualitative change. 

While the single-site DMFT cannot open the Mott gap for any physical value of $U$, the MO-DMFT yields a qualitatively correct result even when applied to the simplest $T^2$-triplet MO as shown in Fig.~\ref{fig:MOT1}(b). Therein the splitting of the $T^2$ states into the lower and upper Hubbard bands can be seen, depicted in red hue in the spectral function plot (and the red curve in the PDOS), which leads to the opening of a charge gap. Note that since the $T^2$ triplet is 1/6-filled, it is not possible to obtain an insulating phase in the band picture without breaking both the cubic and time-reversal symmetries~\cite{Sieberer2007}, while in the Mott phase both symmetries can be kept. Hence we conclude that the cluster-MO description is indeed crucial in describing the Mott physics of GaV$_4$S$_8$, at least in its cubic and paramagnetic phase. Note that a similar result was previously reported on GaTa$_4$Se$_8$ by employing maximally-localized Wannier functions for the $T^2$ triplet and solving the Hubbard model via DMFT~\cite{Camjayi2014}. However, as we will show below, this approach overestimates the tendency toward V$_4$ clustering since it ignores the important effect of the Hund's coupling between the $T^2$ and other MOs  on the structural degrees of freedom.

{\color{NavyBlue}\it $T^2 \oplus E$ subspace and Hund's coupling.} 
Despite the appearance of the Mott phase within the simplest $T^2$-MO-DMFT calculation, this is a crude approximation because other MO states are separated from the $T^2$ manifold by less than a fraction of an eV, and the Coulomb repulsion as well as the Hund's coupling are larger or comparable to this separation. Therefore it is important to check what is the effect of including the next set of orbitals into the correlated space. Recently it was shown that the Hund's coupling can have a very strong effect on the strength of correlations by promoting the local high-spin state and consequently allowing spins to decouple from the orbitals, thus allowing strong orbital differentiation~\cite{Haule_njp,Yin-nm11,powerlaws,Hundreview}. Such physics is completely absent in the $T^2$ model, as we assumed that the $E$ MOs are completely filled and inert, leaving a single electron in the $T^2$ MO set.

We next treat the combination of $T^2 \oplus E$ MOs as our correlated subset. Fig.~\ref{fig:MOT2}(a-c) shows the orbital-projected spectral functions from calculations with $J_{\rm H}$ = 0.5, 1.0, and 1.5 eV, respectively ($T$ = 232 K, $U$ = 8 eV). The red and green colors represent the $T^2$ and $E$ characters respectively. The signature of a low-to-high spin crossover, from the $S$ = 1/2 to 5/2 configuration, can be noticed in the plots where the fully occupied $E$ doublet (at $J_{\rm H}$ = 0.5 eV) begins to lose spectral weight as $J_{\rm H}$ is enhanced. Tracking the Monte Carlo probabilities for the $S_z$ = 1/2 and 5/2 states, plotted in Fig.~\ref{fig:MOT2}(d), shows the same tendency that the $S_z$ = 1/2 probability decreases and collapses almost to zero around $J_{\rm H}$ $\sim$ 1 eV. Note that we report $S_z$ values rather than $S$ values, because of our choice of an Ising-type approximation of the Coulomb interaction in the MO-DMFT impurity solver~\footnote{This approximation leads to some mixing between half-integer spin states, but is not expected to change qualitative aspects of the results}. For $J_{\rm H} \gtrsim$ 1 eV, it can be seen that the $E$ doublet becomes half-filled (see Fig.~\ref{fig:MOT2}(c) and (e)), showing that the crossover to the high-spin state is almost complete. Note that even a moderate $J_{\rm H} \lesssim$ 1 eV, appropriate for 3$d$ transition-metal compounds~\cite{Vaugier2012}, induces substantial mixing between the low-spin and high-spin states. Therefore one may suspect a potential role of the Hund's coupling physics in the high-temperature cubic phase of GaV$_4$S$_8$. Unexpectedly, it turns out that the Hund's coupling significantly weakens the degree of the V$_4$S$_4$ clustering, in contrast with the Coulomb repulsion $U$ which enhances the clustering, as shown in the following. 

\begin{figure} 
  \centering
  \includegraphics[width=0.48\textwidth]{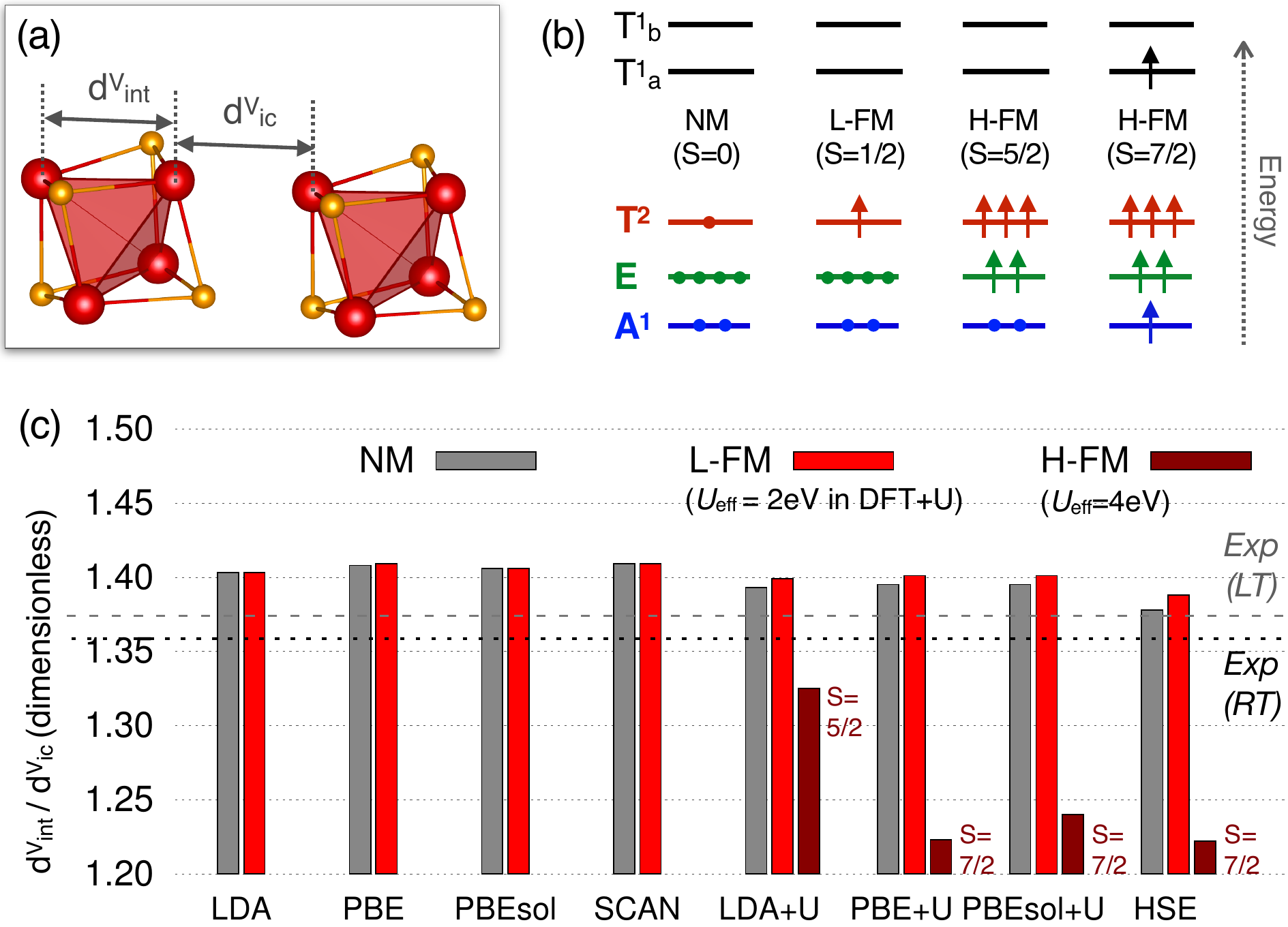}
  \caption{
  (a) Definitions of the intra- and inter-cluster V-V bond lengths $d^{\rm V}_{\rm int}$ and $d^{\rm V}_{\rm ic}$ respectively. 
  (b) Schematic representations of the nonmagnetic (NM), low-spin (L-FM, $S$ = 1/2), and high-spin ferromagnetic (H-FM, $S$ = 5/2 or 7/2) 
  configurations, where the dots and arrows depict nonmagnetic and magnetic electrons respectively. 
  (c) $d^{\rm V}_{\rm int} / d^{\rm V}_{\rm ic}$ from DFT results with different choices of exchange-correlation potentials: LDA~\cite{LDA}, PBE~\cite{PBE}, PBEsol~\cite{PBEsol}, SCAN meta-GGA functional~\cite{SCAN}, DFT+$U$~\cite{Dudarev}, and HSE06 hybrid functional~\cite{HSE06,HSE06-2}. 
  In the DFT+$U$ results, the L-FM and H-FM configurations are obtained by employing $U_{\rm eff}$ = 2 and 4 eV
  in the simplified rotationally-invariant DFT+$U$ scheme~\cite{Dudarev}. Horizontal gray dashed and black dotted lines
  show the values of $d^{\rm V}_{\rm int} / d^{\rm V}_{\rm ic}$ from experimental structures measured at $T$ = 295 and 20 K respectively~\cite{Pocha2000}. 
  }
  \label{fig:dftrx}
\end{figure}

{\color{NavyBlue}\it V$_4$S$_4$ clustering from DFT.}
A parameter quantifying the size of the V$_4$S$_4$ clustering is the ratio between the nearest-neighbor V-V distances, $d^{\rm V}_{\rm int}$/$d^{\rm V}_{\rm ic}$, where $d^{\rm V}_{\rm int}$ and $d^{\rm V}_{\rm ic}$ denote the inter- and intra-cluster V-V distances respectively as shown in Fig.~\ref{fig:dftrx}(a). $d^{\rm V}_{\rm int}$/$d^{\rm V}_{\rm ic}$ is unity in the ideal spinel structure, while in GaV$_4$S$_8$ the value was reported to be 1.35 at $T$ = 295 K and 1.37 at 20 K respectively (see the horizontal dashed/dotted lines in Fig.~\ref{fig:dftrx}(c))~\footnote{Note that at $T$ = 20 K, the compound has a rhombohedral distortion. The value 1.37 is obtained by averaging $d^{\rm V}_{\rm int}$ and $d^{\rm V}_{\rm ic}$ separately and taking the ratio between them.}. 

Fig.~\ref{fig:dftrx}(c) shows the ratios obtained from DFT calculations with different choices of exchange-correlation functionals~\cite{LDA,PBE,PBEsol,SCAN,Dudarev,HSE06,HSE06-2}, which have been reported to yield different values of lattice parameters. Three distinct magnetic configurations were considered: a nonmagnetic configuration (NM), a low-spin ferromagnetic configuration (L-FM) with $S$ = 1/2, and high-spin ferromagnetic configurations (H-FM) with $S$ = 5/2 or 7/2. These are schematically illustrated in Fig.~\ref{fig:dftrx}(b). Note that because the $V_4$ cluster is believed to host a cluster spin moment, FM configurations were considered in our DFT calculations as appropriate for systems with local moments.  

Remarkably, the values of $d^{\rm V}_{\rm int}$/$d^{\rm V}_{\rm ic}$ shown in Fig.~\ref{fig:dftrx}(c) are almost identical, at about 1.4, for all the results on the NM or L-FM configurations, despite different optimized lattice parameters (except HSE, see below). Thus, the degree of clustering is consistently overestimated compared to experimental values.  On the other hand, the H-FM solutions with the DFT+$U$ or HSE06 hybrid functionals severely underestimate the clustering, as shown in Fig.~\ref{fig:dftrx}(c). We notice that in H-FM solutions the lowest occupied MO bonding states ($E$, $A^1$) have been emptied at the expense of occupying higher nonbonding- or antibonding-like states. Therefore it is natural that H-FM solutions show a reduced tendency to clustering. Hence it appears that the small but significant discrepancy between the theoretical (in NM or L-FM) and experimental $d^{\rm V}_{\rm int}$/$d^{\rm V}_{\rm ic}$ values results from the small admixture of the high-spin configurations to the dominant low-spin configuration in the electronic states of GaV$_4$S$_8$, which cannot be captured in the framework of conventional DFT. Note that even though the HSE06 results with NM or L-FM configurations seem to reproduce reasonable $d^{\rm V}_{\rm int}$/$d^{\rm V}_{\rm ic}$ values, those states are much higher in energy by 1.5 eV\,/\,f.u. compared to the $S$ = 7/2 H-FM phase. Also, all of the DFT results (NM, L-FM, and H-FM) fail to reproduce the insulating phase, signifying the failure of the DFT methods in this system.

\begin{figure}
  \centering
  \includegraphics[width=0.48\textwidth]{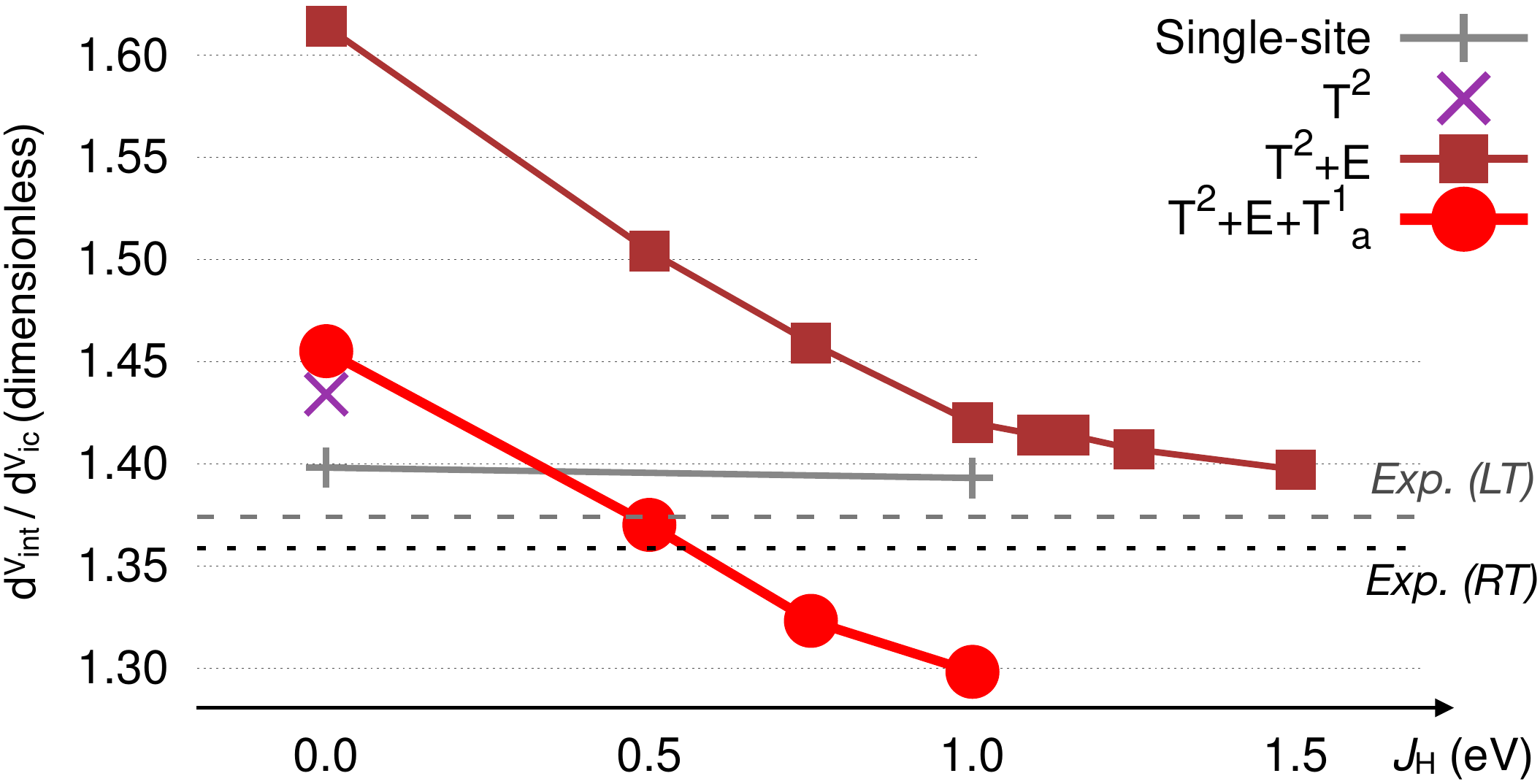}
  \caption{
  $d^{\rm V}_{\rm int} / d^{\rm V}_{\rm ic}$ from DFT results as a function of $J_{\rm H}$. Note that MO-$T^2$ and
  MO-$\{T^2 \oplus T^1_a\}$ configurations are not affected by $J_{\rm H}$ because of the single occupancy, and that the MO-$\{T^2 \oplus E \oplus T^1_a\}$ reaches the experimental $d^{\rm V}_{\rm int} / d^{\rm V}_{\rm ic}$
  near $J_{\rm H}$ = 0.5 eV. 
  }
  \label{fig:dV}
\end{figure}

{\color{NavyBlue}\it V$_4$S$_4$ clustering from MO-DMFT.}
Figure~\ref{fig:dV} shows the evolution of the $d^{\rm V}_{\rm int}$/$d^{\rm V}_{\rm ic}$ values from the DMFT results. As explained above, within the single-site DMFT the correlations appear to be weak, so that the predicted structure is very close to the DFT prediction. As the intra-cluster correlations are considered via the $T^2$ MO, the local Hubbard $U$ enhances the clustering tendency, which is clear from the predicted values at $J_{\rm H}$ = 0. It can be seen that the clustering tendency is substantially overemphasized when the $T^2\oplus E$ are considered as correlated, due to the bonding nature of the $E$ MO. When the antibonding $T^1_a$ MO is also included, the degree of clustering reverts back to similar value as for the $T^2$-only calculation. Still, the value of $d^{\rm V}_{\rm int}$/$d^{\rm V}_{\rm ic}$ is larger than the DFT-optimized one at $J_{\rm H}$ = 0, showing the role of $U$ in enhancing the clustering. 

Once the Hund's coupling is turned on, the degree of clustering is quickly reduced (except for the $T^2$-only case where there is only one electron) as shown in Fig.~\ref{fig:dV}. We then obtain the experimental $d^{\rm V}_{\rm int}$/$d^{\rm V}_{\rm ic}$ values around $J_{\rm H}$ = 0.5 eV, which is a reasonable value for our model, in which $e_{\rm g}$ states (as well as $A^1$ and $T^1_b$) are screening the interaction. This observation is consistent with the spectroscopic tendency mentioned above, where $J_{\rm H}$ promotes the high-spin state so that spin moments can be more localized on each V site. We thus find, quite surprisingly, that in cases with strong clustering the Coulomb $U$ and Hund's $J_{\rm H}$ can play opposite roles: the former promotes non-local correlations and formation of the bonding molecular orbital state, while the latter promotes local atom-centered high-spin states. This Janus-faced effect of $U$ and $J_{\rm H}$ is a central result of this study. Note also that the reduction of $d^{\rm V}_{\rm int}$/$d^{\rm V}_{\rm ic}$ is significant already at $J_{\rm H}$ = 0.5 eV, where the mixture of the high-spin configurations is quite small as shown in Fig.~\ref{fig:MOT2}(d). This implies an unusual strong coupling between the electronic configuration and the V$_4$ clustering, which may be exploited to tune the spin configuration by employing optical pumping techniques as done in VO$_2$~\cite{Zheng2011ARMR}. 

{\color{NavyBlue}\it Discussion and Summary.}
Im summary, in this work we have clarified the significance of electron correlations in describing the MO Mott physics and structural properties of GaV$_4$S$_8$, especially the Janus-faced role of $U$ and $J_{\rm H}$ in its crystal structure, which can be extended to study the low-temperature ferroelectric and multiferroic phases~\cite{Kezsmarki2015,Ruffe1500916,Widmann2017} of the same compound and possible unconventional electron-lattice couplings therein. With a careful choice of the MO correlated subspace, our MO-based DMFT approach can tackle systems with large-sized clusters that are not amenable to solution using conventional cluster DMFT approaches, such as $1T$-Ta\{S,Se\}$_2$ and other cluster Mott insulating systems~\cite{Gang1,Gang2}.
 
\begin{acknowledgments}
{\color{NavyBlue}\it Acknowledgments}:
This work was supported by NSF DMREF DMR-1629059. HSK was funded by the National Research Foundation of Korea (Basic Science Research Program, Grant No. 2020R1C1C1005900), and also thanks the National Supercomputing Center of Korea for the support of supercomputing resources including technical assistances (Grant No. KSC-2019-CRE-0036).   
\end{acknowledgments}

\bibliography{GVS_CDMFT}{}

\newpage
\onecolumngrid

\renewcommand{\thefigure}{S\arabic{figure}}

\subsection{Density functional theory calculations}

For unit cell optimizations (cell volume and shape) and relaxations of initial internal coordinates, the Vienna {\it ab-initio} Simulation Package ({\sc vasp}), which employs the projector-augmented wave (PAW) basis set~\cite{VASP1,VASP2}, was used for density functional theory (DFT) calculations in this work. 330 eV of plane-wave energy cutoff (PREC=high) and 15$\times$15$\times$15 $\Gamma$-centered $k$-grid sampling were employed. For the treatment of electron correlations within DFT, several exchange-correlation functional were employed, including Ceperley-Alder (CA) parametrization of local density approximation~\cite{LDA}, Perdew-Burke-Ernzerhof generalized gradient approximation (PBE)~\cite{PBE} and its revision for crystalline solids (PBEsol)~\cite{PBEsol}, SCAN meta-GGA functional~\cite{SCAN}, DFT+$U$~\cite{Dudarev} on top of LDA, PBE, and PBEsol, and HSE06 hybrid functional~\cite{HSE06,HSE06-2}. $10^{-4}$ eV/\AA~of force criterion was employed for structural optimizations. 

\subsection{Cluster dynamical mean-field theory calculations}
A fully charge-self-consistent dynamical mean-field method\cite{Dmft}, implemented in DFT + Embedded DMFT (eDMFT) Functional code (\href{http://hauleweb.rutgers.edu/tutorials/}{http://hauleweb.rutgers.edu/tutorials/}) which is combined with {\sc wien2k} code\cite{wien2k}, is employed for computations of electronic properties and optimizations of internal coordinates\cite{Force2016}. At the DFT level the Perdew-Wang (PW) local density approximation is employed, which was argued to yield the best agreement of lattice properties when combined with DMFT\cite{Haule2015FE}. 15$\times$15$\times$15 $\Gamma$-centered $k$-grid was used to sample the first Brillouin zone with $RK_{\rm max}$ = 7.0. A force criterion of 10$^{-4}$ Ry/Bohr was adopted for optimizations of internal coordinates. The cubic lattice parameter was fixed to be the experimental value reported in Ref.~\onlinecite{Pocha2000}. 

A continuous-time quantum Monte Carlo method in the hybridization-expansion limit (CT-HYB) was used to solve the auxiliary quantum impurity problem\cite{HauleQMC}. For the CT-HYB calculations, up to $3 \times 10^{10}$ Monte Carlo steps were employed for each Monte Carlo run. In most runs temperature was set to be 232K, but in calculations with 8 molecular orbitals (MOs) ($T^2 \oplus E \oplus T^1_a$ in Fig. 1 in the main text) as the correlated subspace it was increased up to 1160K because of the increased computational cost. -10 to +10 eV of hybridization window (with respect to the Fermi level) was chosen, and the on-site Coulomb interaction parameters $U$ and $J_{\rm H}$ for V $t_{\rm 2g}$ orbitals were varied within the range of 6 $\sim$ 8 eV and 0 $\sim$ 1.5 eV, respectively. A simplified Ising-type (density-density terms only) Coulomb interaction was employed in this work, and it was tested that the use of full Coulomb interaction yields only quantitative difference in results with MO-$T^2$ and $T^2\oplus E$ (not tested for MO-$T^2\oplus E \oplus T^1_a$ case due to the high cost, see Sec. \ref{SI_full_vs_Ising}). A nominal double counting scheme was used, with the MO occupations for double counting corrections for the V$_4$ cluster were chosen to be 1 or 5, depending on the choice of correlated subspace; 1 for MO-$T^2$ and $T^2\oplus T^1_a$, and 5 for other cases with including $E$ in the correlated subspace. 

In the CT-HYB calculations of the $T^2 \oplus E \oplus T^1_a$ MO subspace, MO multiplet states with the occupancy $n\leq 7$ were kept (26,333 states out of $4^8$ = 65,536 states in the 8 orbital Fock space) to reduce the computational cost, where the average impurity occupancy was $\sim$ 5. It was checked that the sum of probabilities for $n\geq 8$ configurations are less than 1 percent. The high-frequency tail of the Green's function was calculated via the Hubbard-I approximation. 

We comment that, due to the quite small intercluster hybridization, the perturbation order is small in our CT-HYB formalism, with the average perturbation order being less than 80 for the case of the largest correlated subspace ($T^2 \oplus E \oplus T^1_a$). In addition, it is shown below that the cubic symmetry of the V$_4$ cluster enforces the form of Coulomb interaction matrix between the molecular orbitals to be identical to that of atomic orbitals, at least for the $T^2$ molecular orbital. Hence the negative sign problem in our CT-HYB formalism is suppressed, which greatly facilitates the computation in addition to the small perturbation order. 

For the computational resources, we used 8 Intel Xeon E5-2680 v4 CPUs (2.4 GHz, total 112 CPU cores) and about 400GB of memory for the case of the largest correlated subspace ($T^2 \oplus E \oplus T^1_a$). Even with the use of a high temperature $T$ = 1160K and truncation of superstates in the CT-HYB stage, employing the full Coulomb interaction or a lower temperature such as $T$ = 780K requires allocation of a memory size that exceeds the limitation of our hardware (512 GB). For the continuation of this study, either larger computational resources or more efficient ways to treat the less-occupied $T^1_a$ orbital would be necessary.

As for possible inter-site, intra-cluster Coulomb repulsion terms, the screened
Coulomb repulsion in solid state compounds can be fitted to a Yukawa-like form $ V(r) \simeq e^{-\lambda r}/\epsilon r$, and by comparing the unscreened and screened Coulomb parameters ($U$ and $J$), one can obtain the screening length $\lambda$ and electric permeability $\epsilon$ for a given system~\cite{Haule_exactDC}. The values of unscreened $U$ and $J$ can be directly computed by using the local orbital projectors, and it was shown in a recent study that reasonable values of the screened $U$ and $J$ for the 3$d$ transition metal elements in our DFT+DMFT implementation are 10 and 1 eV respectively~\cite{Birol2014}. With these we get $\lambda \simeq 0.52$ and $\epsilon \simeq 1.09$, which yields $\int d{\bf r} \int d{\bf r}' \rho({\bf r}) \rho({\bf r}') \frac{e^{-\lambda \vert {\bf r} - {\bf r}' - {\bf R} \vert}}{\epsilon \vert {\bf r} - {\bf r}' - {\bf R} \vert} \simeq 0.15$ eV ($\vert {\bf R} \vert$ being the intra-cluster V-V distance). This value is an order of magnitude smaller than the intra-cluster hybridization and on-site Coulomb interaction, so we conclude that the effect of intra-cluster, inter-site Coulomb repulsion is insignificant. 

\subsection{Projecting the on-site Coulomb interactions onto the MO subspace}

Note that the $U$ and $J_{\rm H}$ are parameters defined for the atomic orbitals, which should be unitary transformed and projected onto the MOs for the impurity solver. More generally, the Coulomb repulsion matrix elements $U_{m_1, m_2, m'_1, m'_2}$ at an atomic site have the form,
\begin{align}
U_{m_1, m_2, m'_1, m'_2} &= 
	\sum_{m,k} \frac{4}{2\pi+1}
		\langle Y_{lm_1} \vert Y_{km} \vert Y_{lm'_1} \rangle
		\langle Y_{lm_2} \vert Y^*_{km} \vert Y_{lm'_2} \rangle F^k,
\end{align}
where $F^k$ are nonzero only for $k$ = 0, 2, 4 for $d$-orbitals ($l$ = 2) and $\langle Y_{lm_1} \vert Y_{km} \vert Y_{lm'_1} \rangle$ are Clebsch-Gordan coefficients. We introduce the MO states
\begin{align}
\vert D_\alpha \rangle &= \sum_{im} (Q^\dag)^{im}_\alpha \vert Y^i_{lm} \rangle,
\end{align}
where $Q$ is the unitary transform between the MO and the atomic orbitals, and $\alpha$ and $i=1,\cdots,4$ are the MO orbital and atomic site indices respectively. Then the Coulomb repulsion matrix elements for the MO states $U_{\alpha_1, \alpha_2, \alpha'_1, \alpha'_2}$ can be written as
\begin{align}
U_{\alpha_1, \alpha_2, \alpha'_1, \alpha'_2} &= 
	\sum_{i,m,k} \frac{4}{2\pi+1}
		\langle D_{\alpha_1} \vert Y^i_{km} \vert D_{\alpha'_1} \rangle
		\langle D_{\alpha_2} \vert Y^{i*}_{km} \vert D_{\alpha'_2} \rangle F^k \\
		&\sim (QQQ^\dag Q^\dag)^{i\{m\}}_{\{\alpha\}} U^i_{\{m\}}. 
\end{align}
Note that the inter-site Coulomb interactions were ignored here, which can be considered insignificant in 3$d$ transition metal compounds. 

Below we show explicitly how the on-site Coulomb interactions projected onto the $T^2$ triplet subspace should look like. As shown in Fig.~1 in the main text, electronic structure near the Fermi level ([-1eV, 1eV] window with respect to the Fermi level) is dominated by the 
atomic $t_{\rm 2g}$ orbitals of V due to the distorted but prevalent cubic VS$_6$ octahedral environment. Therefore choosing 
12 $t_{\rm 2g}$ orbitals as our main interest is a reasonable choice. For simplicity we chose the Kanamori form of the Coulomb interaction, which is written in a normal-ordered form as follows;
\begin{align}
\hat{H}_K = -\sum_i \Big[
(U-2J) & \sum_{mm'} \hat{d}^\dag_{im\uparrow} \hat{d}^\dag_{im'\downarrow} \hat{d}_{im\uparrow} \hat{d}_{im'\downarrow} \nonumber \\
+ 2J & \sum_{m} \hat{d}^\dag_{im\uparrow} \hat{d}^\dag_{im\downarrow} \hat{d}_{im\uparrow} \hat{d}_{im\downarrow} \nonumber \\
+ \frac{U-3J}{2} & \sum_{m \neq m',\sigma}
\hat{d}^\dag_{im\sigma} \hat{d}^\dag_{im'\sigma} \hat{d}_{im\sigma} \hat{d}_{im'\sigma} \nonumber \\
-J & \sum_{m \neq m'} \hat{d}^\dag_{im\uparrow} \hat{d}^\dag_{im'\downarrow} \hat{d}_{im\downarrow} \hat{d}_{im'\uparrow} \nonumber \\
-J & \sum_{m \neq m'} \hat{d}^\dag_{im\uparrow} \hat{d}^\dag_{im\downarrow} \hat{d}_{im'\downarrow} \hat{d}_{im'\uparrow}
\Big].
\end{align}
Here $i$, $\sigma$, and $m$, $m'$ are site, spin, and orbital indices for Cartesian $t_{\rm 2g}$ 
orbitals ($d_{xz,yz,xy}$) respectively. 

Now we introduce the MO creation/annihilation operators;
\begin{align}
\hat{d}_{im\sigma} &= \sum_{\alpha} Q^\alpha_{im} \hat{D}_{\alpha\sigma} \\
\hat{d}^\dag_{im\sigma} &= \sum_{\alpha} (Q^\dag)^{im}_\alpha \hat{D}^\dag_{\alpha\sigma}
\end{align}
where $\alpha$ runs over the 12 molecular orbitals and we are ignoring spin-orbit coupling (SOC) at this stage. 
$Q^\alpha_{im}$ is the 12$\times$12 transformation matrix from the atomic $t_{\rm 2g}$ to the MO spaces. 
In terms of {\it global} coordinates (using the same cartesian coordinates for all $V$ sites) it is tabulated in 
Table \ref{tab:Q}. Note that in actual calculations, since the four V sites are equivalent to each other up to a symmetry 
operation, $Q$ should be unitarily transformed to a local coordinate system at each V site. 

\begin{table}
\centering
\begin{tabular}{llrrr|rrr|rrr|rrr}  \hline\hline
Irreps & No. & \multicolumn{12}{l}{Coeff.} \\
&& \multicolumn{3}{l}{V$_1$ (0.4,0.4,0.4)}& \multicolumn{3}{l}{V$_2$ (0.4,0.6,0.6)} &
\multicolumn{3}{l}{V$_3$ (0.6,0.6,0.4)} & \multicolumn{3}{l}{V$_4$ (0.6,0.4,0.6)} \\
&& $d_{xy}$ & $d_{yz}$ & $d_{xz}$ & $d_{xy}$ & $d_{yz}$ & $d_{xz}$ 
& $d_{xy}$ & $d_{yz}$ & $d_{xz}$ & $d_{xy}$ & $d_{yz}$ & $d_{xz}$ \\ \hline
$A$ & 1 &   +1 & +1 & +1 &  -1 & +1 & -1 &  +1 & -1 & -1&  -1 & -1 & +1 \\[7pt]
$E$ & 1 &   +1 & +$w^1$ & +$w^2$ &   -1 & +$w^1$ & -$w^2$ &   +1 & -$w^1$ & -$w^2$ &   -1 & -$w^1$ & +$w^2$ \\
       & 2 &   +1 & +$w^2$ & +$w^1$ &   -1 & +$w^2$ & -$w^1$ &   +1 & -$w^2$ & -$w^1$ &   -1 & -$w^2$ & +$w^1$ \\[7pt]
$T^2$ & 1 &   +1 & 0 & 0 &   +1 & 0 & 0 &   +1 & 0 & 0 &   +1 & 0 & 0 \\
           & 2 &   0 & +1 & 0 &   0 & +1 & 0 &   0 & +1 & 0 &   0 & +1 & 0 \\
           & 3 &   0 & 0 & +1 &   0 & 0 & +1 &   0 & 0 & +1 &   0 & 0 & +1 \\[7pt]
$T^1_a$ & 1 &   0 & +1 & -1 &   0 & -1 & -1 &   0 & -1 & +1 &   0 & +1 & +1 \\
           & 2 & +1 & 0 & -1 &   -1 & 0 & +1 &  -1 & 0 & -1 &  +1 & 0 & +1 \\
           & 3 & +1 & -1 & 0 &  +1 & +1 & 0 &  -1 & -1 & 0 &  -1 & +1 & 0 \\[7pt]
$T^1_b$ & 1 &   0 & +1 & +1 &   0 & -1 & +1 &   0 & -1 & -1 &   0 & +1 & -1 \\
           & 2 & +1 & 0 & +1 &   -1 & 0 & -1 &  -1 & 0 & +1 &  +1 & 0 & -1 \\
           & 3 & +1 & +1 & 0 &  +1 & -1 & 0 &  -1 & +1 & 0 &  -1 & -1 & 0 \\
 \hline\hline
\end{tabular}
\caption{
Transformation matrix $Q^\alpha_{im}$ from atomic $t_{\rm 2g}$ to molecular orbital basis before normalization,
where $w=e^{2\pi i/3}$. 
}
\label{tab:Q}
\end{table}

Plugging them into $\hat{H}_{K}$ yields, 
\begin{align}
\hat{H}_K = -\sum_{\alpha\beta\gamma\delta} \Big[
(U-2J)& \sum_{i} \left\{ \sum_{mm'} 
		(Q^\dag)^{im}_\alpha (Q^\dag)^{im'}_\beta  Q^\gamma_{im} Q^\delta_{im'} \right\}
		\hat{D}^\dag_{\alpha\uparrow} \hat{D}^\dag_{\beta\downarrow} \hat{D}_{\gamma\uparrow} \hat{D}_{\delta\downarrow} \nonumber \\
+2J & \sum_{i} \left\{ \sum_{m} 
		(Q^\dag)^{im}_\alpha (Q^\dag)^{im}_\beta  Q^\gamma_{im} Q^\delta_{im} \right\}
		\hat{D}^\dag_{\alpha\uparrow} \hat{D}^\dag_{\beta\downarrow} \hat{D}_{\gamma\uparrow} \hat{D}_{\delta\downarrow} \nonumber \\
+\frac{U-3J}{2} & \sum_{i} \left\{ \sum_{m\neq m'} 
		(Q^\dag)^{im}_\alpha (Q^\dag)^{im'}_\beta  Q^\gamma_{im} Q^\delta_{im'} \right\}
		\sum_\sigma \hat{D}^\dag_{\alpha\sigma} \hat{D}^\dag_{\beta\sigma} \hat{D}_{\gamma\sigma} \hat{D}_{\delta\sigma} \nonumber \\
-J & \sum_{i} \left\{ \sum_{m\neq m'} 
		(Q^\dag)^{im}_\alpha (Q^\dag)^{im'}_\beta  Q^\gamma_{im} Q^\delta_{im'} \right\}
		\hat{D}^\dag_{\alpha\uparrow} \hat{D}^\dag_{\beta\downarrow} \hat{D}_{\gamma\downarrow} \hat{D}_{\delta\uparrow} \nonumber \\
-J & \sum_{i} \left\{ \sum_{m\neq m'} 
		(Q^\dag)^{im}_\alpha (Q^\dag)^{im}_\beta  Q^\gamma_{im'} Q^\delta_{im'} \right\}
		\hat{D}^\dag_{\alpha\uparrow} \hat{D}^\dag_{\beta\downarrow} \hat{D}_{\gamma\downarrow} \hat{D}_{\delta\uparrow} 
\Big].
\end{align}
In the above expression, product of $Q$s can be rewritten as
\begin{align}
\left( Q^\dag \otimes Q^\dag \right)^{imm'}_{\alpha\beta} &\equiv (Q^\dag)^{im}_\alpha (Q^\dag)^{im'}_\beta \\
\left( Q \otimes Q \right)_{imm'}^{\gamma\delta} &\equiv Q_{im}^\gamma Q_{im'}^\delta,
\end{align}
and, since we are considering {\it local} Coulomb interactions, we are taking direct products of $i$-subsections ($i$=1,$\cdots$,4) of
$Q$ and $Q^\dag$ matrices, so that $Q \otimes Q$ (and $Q^\dag \otimes Q^\dag$) 
has dimension of 9$\times$144 for each $i$ when we are considering the full 12-dimensional molecular orbital space. \\

Since we don't include SOC and the transformation matrices does not have spin indices, 
all $(Q^\dag \otimes Q^\dag) \cdot (Q \otimes Q)$ terms are free of spin components and can be classified into 
four different kinds; i) $\sum_{mm'} \left( Q^\dag \otimes Q^\dag \right)^{imm'}_{\alpha\beta} \left( Q \otimes Q \right)_{imm'}^{\gamma\delta}$,
ii) $\sum_{m} \left( Q^\dag \otimes Q^\dag \right)^{imm}_{\alpha\beta} \left( Q \otimes Q \right)_{imm}^{\gamma\delta}$,
iii) $\sum_{m \neq m'} \left( Q^\dag \otimes Q^\dag \right)^{imm'}_{\alpha\beta} \left( Q \otimes Q \right)_{imm'}^{\gamma\delta}$,
and iv) $\sum_{m \neq m'} \left( Q^\dag \otimes Q^\dag \right)^{imm}_{\alpha\beta} \left( Q \otimes Q \right)_{im'm'}^{\gamma\delta}$. 
Here case iii) is just the subtraction of ii) from i). \\

Computation of the transformation matrix is straightforward, but now all different molecular orbitals can mix 
even in a simple density-density interaction form (the first three terms in $\hat{H}_K$). However, 
things become much simpler in the most basic case of 
considering only the $T^2$ irrep as the correlated subspace. In that case, all $Q_i$ (and $Q^{\dag,i}$) become 3$\times$3
identity matrix (with normalization factor 1/2), so that all $Q \otimes Q$ and $Q^\dag \otimes Q^\dag$ become 9$\times$9
identity matrix with a prefactor 1/4, so that 
\begin{align}
{\rm i)} & \sum_{imm'} \left( Q^\dag \otimes Q^\dag \right)^{imm'}_{\alpha\beta} \left( Q \otimes Q \right)_{imm'}^{\gamma\delta}
	 \rightarrow \frac{1}{4} \delta_{\alpha\gamma} \delta_{\beta\delta}, \\
{\rm ii)} & \sum_{im} \left( Q^\dag \otimes Q^\dag \right)^{imm}_{\alpha\beta} \left( Q \otimes Q \right)_{imm}^{\gamma\delta}
	 \rightarrow \frac{1}{4} \delta_{\alpha\gamma} \delta_{\beta\delta}  \delta_{\alpha\beta}, \\
{\rm iii)} & \sum_{i,m \neq m'} \left( Q^\dag \otimes Q^\dag \right)^{imm'}_{\alpha\beta} \left( Q \otimes Q \right)_{imm'}^{\gamma\delta}
	 \rightarrow \frac{1}{4} \delta_{\alpha\gamma} \delta_{\beta\delta} (1-\delta_{\alpha\beta}), \\
{\rm iv)} & \sum_{m \neq m'} \left( Q^\dag \otimes Q^\dag \right)^{imm}_{\alpha\beta} \left( Q \otimes Q \right)_{im'm'}^{\gamma\delta}
	 \rightarrow \frac{1}{4} \delta_{\alpha\beta} \delta_{\gamma\delta} (1-\delta_{\alpha\gamma}).
\end{align}
Hence $\hat{H}_{K}$, projected onto the MO-$T^2$ subspace, becomes
\begin{align}
\hat{H}^{\rm MO}_K = -\frac{1}{4}\Big[
(U-2J) & \sum_{mm'} \hat{D}^\dag_{m\uparrow} \hat{D}^\dag_{m'\downarrow} \hat{D}_{m\uparrow} \hat{D}_{m'\downarrow} \nonumber \\
+ 2J & \sum_{m} \hat{D}^\dag_{m\uparrow} \hat{D}^\dag_{m\downarrow} \hat{D}_{m\uparrow} \hat{D}_{m\downarrow} \nonumber \\
+ \frac{U-3J}{2} & \sum_{m \neq m',\sigma}
\hat{D}^\dag_{m\sigma} \hat{D}^\dag_{m'\sigma} \hat{D}_{m\sigma} \hat{D}_{m'\sigma} \nonumber \\
-J & \sum_{m \neq m'} \hat{D}^\dag_{m\uparrow} \hat{D}^\dag_{m'\downarrow} \hat{D}_{m\downarrow} \hat{D}_{m'\uparrow} \nonumber \\
-J & \sum_{m \neq m'} \hat{D}^\dag_{m\uparrow} \hat{D}^\dag_{m\downarrow} \hat{D}_{m'\downarrow} \hat{D}_{m'\uparrow}
\Big].
\label{eq:HK}
\end{align}

Note that $\hat{H}^{\rm MO}_K$ has the exactly same form with the atomic $\hat{H}_K$, except the prefactor 1/4 because of the equidistribution of the MO-$T^2$ wavefunctions all over the four V sites. 

\subsection{Choice of Coulomb interactions and V$_4$ clustering}
\label{SI_full_vs_Ising}

\begin{figure}
  \centering
  \includegraphics[width=0.6\textwidth]{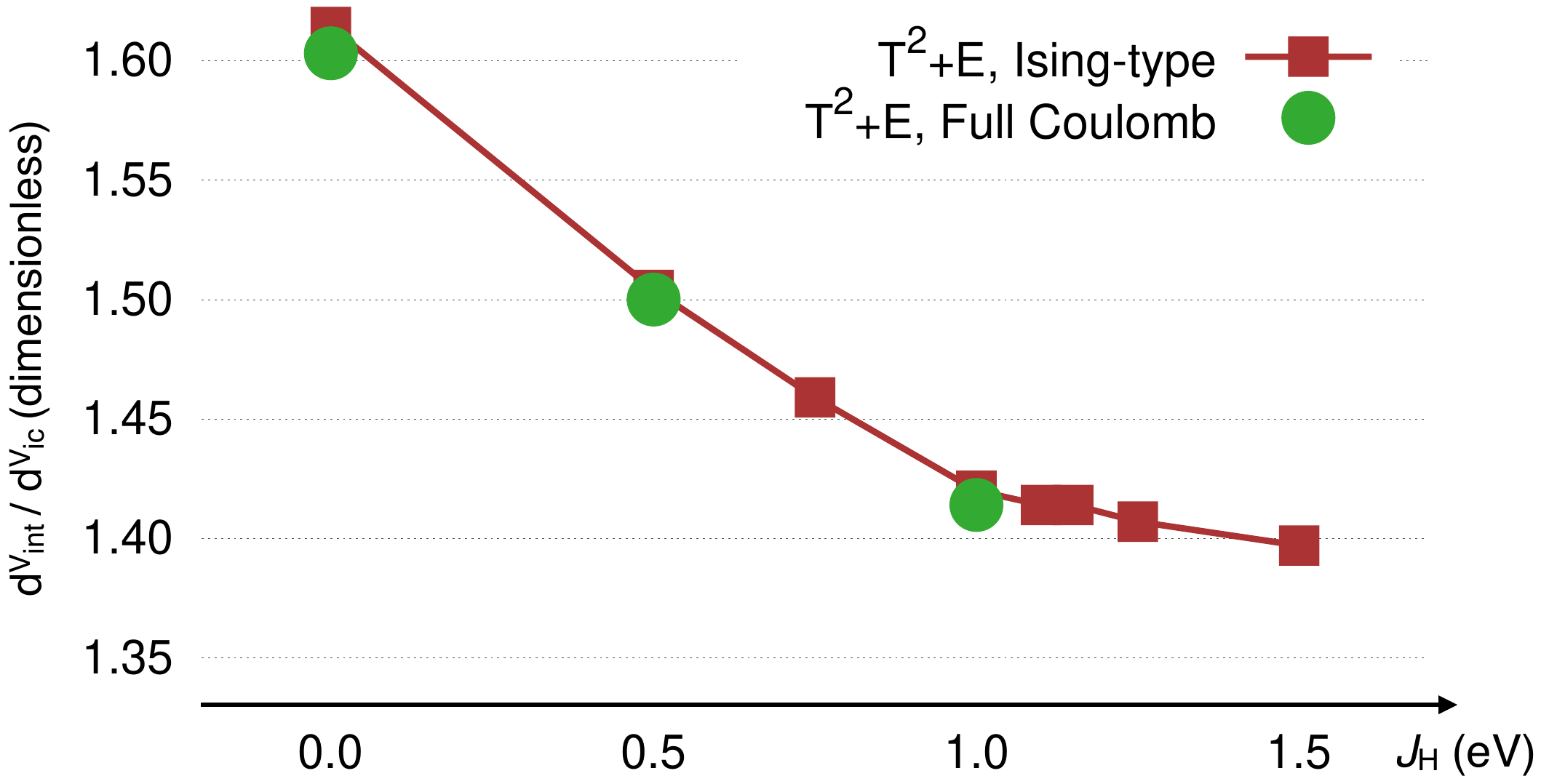}
  \caption{
 $d^{\rm V}_{\rm int} / d^{\rm V}_{\rm ic}$ from different choices (Ising-like and full Coulomb) of on-site Coulomb interactions as a function of $J_{\rm H}$, where MO-$\{T^2 \oplus E\}$ is employed as the correlated orbitals. Note that two results show similar $d^{\rm V}_{\rm int} / d^{\rm V}_{\rm ic}$ values and same tendency with respect to the increasing $J_{\rm H}$. 
}
  \label{figS:dV}
\end{figure}

To check the reliability of employing Ising-like (density-density type) Coulomb interactions in our study, we compare our results presented in the main text with those employing full Coulomb interactions. For the comparison we chose the MO-$\{T^2 \oplus E\}$ as our correlated subspace because computational costs using full Coulomb interactions in the MO-$\{T^2 \oplus E \oplus T^1_a\}$ configuration exceeds our hardware limit. We would like to argue that, if the tendency of $d^{\rm V}_{\rm int} / d^{\rm V}_{\rm ic}$ as a function of $J_{\rm H}$ is consistent across both choices of Coulomb interactions in the MO-$\{T^2 \oplus E\}$ configuration, then it should be so in the MO-$\{T^2 \oplus E \oplus T^1_a\}$ setup as well. This is because effects of Hund's coupling are most dominant within the MO-$\{T^2 \oplus E\}$ subspace, and although the inclusion of the $T^1_a$ orbital is crucial in obtaining realistic value of $d^{\rm V}_{\rm int} / d^{\rm V}_{\rm ic}$, electron occupation in the $T^1_a$ orbital remains small ($<$ 0.1) even in the case of $J_{\rm H} > 1$ eV. 

Figure \ref{figS:dV} shows the comparison of calculated $d^{\rm V}_{\rm int} / d^{\rm V}_{\rm ic}$ between two choices of Coulomb interactions: Ising-like and full. It can be seen that the choice of Coulomb interactions does not make any qualitative differences. While the choice of Ising-like interactions breaks rotational symmetry in the magnetic sector and may affect magnetic properties and metal-insulator transition behaviors, its effects on structural degrees of freedom in our case seems less significant.

\subsection{On-site and inter-site self-energies}

In this section the role of the Hund's coupling is discussed in terms of the real space representation of the self-energy. Here we focus on the $T^2 \oplus E$ subspaces and their self-energies. Similar analysis can be done with other MO subspaces, however, for the purpose of discussing the role of $J_{\rm H}$ it seems that $T^2 \oplus E$ should suffice. 

\begin{figure}
  \centering
  \includegraphics[width=0.6\textwidth]{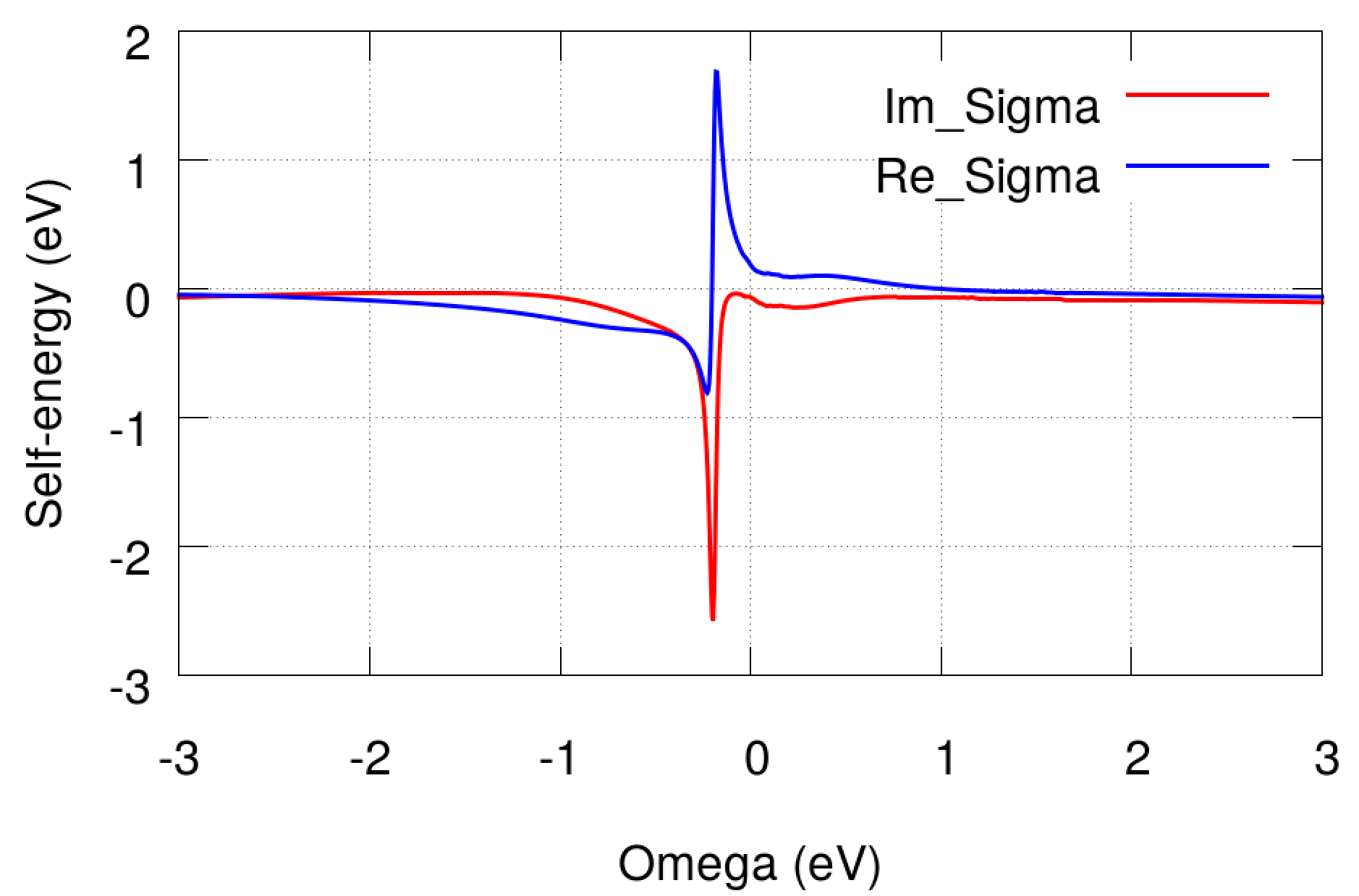}
  \caption{
Real (blue) and imaginary (red) part of ${\Sigma}^{T^2}(\omega)$ after the analytic continuation from the imaginary frequency space.
  }
  \label{figS:T2self}
\end{figure}

In our calculations the cluster self-energies are diagonalized within the MO representation. When back-transformed into the atomic orbital basis representation, on-site (local) and inter-site (non-local) self-energies within the V$_4$ tetrahedron can be obtained. 
In the simplest case with the correlated MO-$T^2$ triplet only, the form of the self-energy in the atomic representation becomes simple; Namely, in the four-site real-space representation (four sites $\otimes$ atomic $t_{\rm 2g}$), all the on-site and inter-site blocks are enforced to be identical due to the choice of the $T^2$ correlated orbitals when the cubic and time-reversal symmetries are present, so that
\begin{align}
\boldsymbol{\Sigma}\left[ T^2 \right] (\omega) \equiv
\frac{1}{4}
\hat{\Sigma}^{T^2} (\omega) \left(
\begin{array}{cccc}
1 & 1 &1 &1 \\
1 & 1 &1 &1 \\
1 & 1 &1 &1 \\
1 & 1 &1 &1
\end{array} \right),
\label{eq:self1}
\end{align}
where each $3 \times 3$ block $\hat{\Sigma}^{T^2} = \Sigma^{T^2} \times \hat{I}_{3\times 3}$ in the atomic $t_{\rm 2g}$ space ($d_{xy}$, $d_{yz}$, and $d_{xz}$), $\hat{I}_{3\times 3}$ is an identity matrix of dimension 3, and the frequency $\omega$ can be either real or imaginary. Note that $\Sigma^{T^2}$ is the diagonal self-energy in the $T^2$-MO representation, and that the prefactor $\frac{1}{4}$ in Eq.~(\ref{eq:self1}) is the one appearing in Eq.~(\ref{eq:HK}). Here we choose the same global coordinate in defining the $t_{\rm 2g}$ orbitals at all sites, and proper coordinate transforms should be applied to each block when represented in local coordinates ($\hat{\Sigma}^{T^2}_{ij} \rightarrow (Q^\dag)_{gi} \hat{\Sigma}^{T^2}_{ij} Q_{jg}$, where the transformation $Q_{ig}$ is made from the global to the site-$i$ local coordinates). Fig.~\ref{figS:T2self} plots the real and imaginary parts of ${\Sigma}^{T^2}$ in the real frequency space, showing a pole in the imaginary part inside the Mott gap.

From this real-space representation of the self-energy, the implication of choosing only the $T^2$ MO as correlated orbitals becomes clearer; i) it introduces the inter-site self-energy in addition to the on-site counterpart, and ii) it prevents the correlations from becoming more local by enforcing the on-site and inter-site self-energies to be identical. The latter, especially, can be a serious issue when the size of the correlations that favor the formation of the local moments, {\it e.g.} the Hund's coupling, becomes comparable to that of inter-site hopping. 

Next, the form of self-energy in the $T^2\oplus E$ is as presented below:
\begin{align}
\boldsymbol{\Sigma}\left[ T^2 \oplus E \right] (\omega)  &= \boldsymbol{\Sigma}\left[ T^2 \right] (\omega)  + \boldsymbol{\Sigma}\left[ E \right] (\omega),
\end{align}
where the $T^2$-part of the self-energy is shown in Eq.~(\ref{eq:self1}). $\boldsymbol{\Sigma}\left[ E \right] (\omega)$ is as follows;
\begin{align}
\boldsymbol{\Sigma}\left[ E \right] (\omega) &\equiv
\left(
\begin{array}{cccc}
\hat{\Sigma}_{11} & \hat{\Sigma}_{12} &\hat{\Sigma}_{13} &\hat{\Sigma}_{14} \\
\hat{\Sigma}^T_{12} & \hat{\Sigma}_{22} &\hat{\Sigma}_{23} &\hat{\Sigma}_{24} \\
\hat{\Sigma}^T_{13} & \hat{\Sigma}^T_{23} &\hat{\Sigma}_{33} &\hat{\Sigma}_{34} \\
\hat{\Sigma}^T_{14} & \hat{\Sigma}^T_{24} &\hat{\Sigma}^T_{34} &\hat{\Sigma}_{44}
\end{array} \right).
\label{eq:self2}
\end{align}
Here the on-site parts $\hat{\Sigma}_{ii}$ are
\begin{align}
\hat{\Sigma}_{ii} &\equiv \Sigma^E (\omega) \left( \frac{1}{6} \hat{I}_{3\times 3} + \frac{1}{12} \hat{\Delta}_{ii} \right),
\end{align}
where $\Sigma^E (\omega)$ is the self-energy for the $E$ doublet in the MO representation, and $\hat{\Delta}_{ii}$ determines the direction of the ``{\it trigonal crystal fields}" to $t_{\rm 2g}$ orbitals at each V site, exerted by $\frac{1}{12}\Sigma^E (\omega) \hat{\Delta}_{ii}$. Namely, if the VS$_6$ octahedron surrounding site 1 is trigonally distorted along the cubic [111] direction with respect to the global Cartesian coordinate ({\it i.e.}, if the site 1 and the center of the V$_4$ cluster are on the same [111] line), then 
\begin{align}
\hat{\Delta}_{11} &= \left(
\begin{array}{ccc}
0 & -1 & -1 \\
-1 & 0 & -1 \\
-1 & -1 & 0
\end{array} \right). 
\end{align}
Other $\hat{\Delta}_{ii}$, for a coordinate choice, should be as follows,
\begin{align}
\hat{\Delta}_{22} &= \left(
\begin{array}{ccc}
0 & +1 & -1 \\
+1 & 0 & +1 \\
-1 & +1 & 0
\end{array} \right),
\hat{\Delta}_{33} = \left(
\begin{array}{ccc}
0 & +1 & +1 \\
+1 & 0 & -1 \\
+1 & -1 & 0
\end{array} \right),
\hat{\Delta}_{44} = \left(
\begin{array}{ccc}
0 & -1 & +1 \\
-1 & 0 & +1 \\
+1 & +1 & 0
\end{array} \right).
\end{align}
Note that this is the coordinate choice that was adopted in this work.

\begin{figure}
  \centering
  \includegraphics[width=1.0\textwidth]{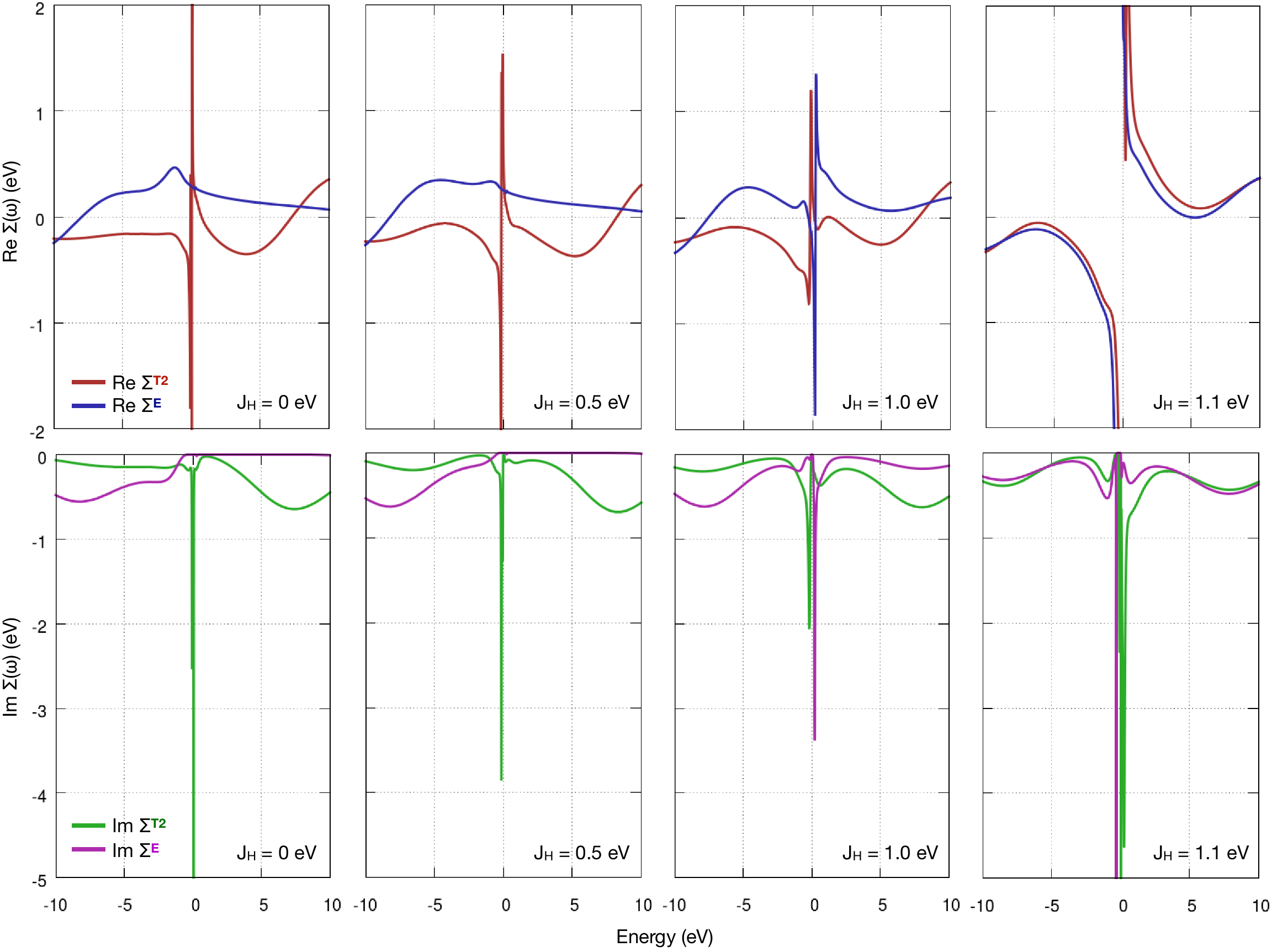}
  \caption{
Real and imaginary part of ${\Sigma}^{T^2,E}(\omega)$ after the analytic continuation from the imaginary frequency space. Top and bottom panels depict real and imaginary parts, respectively. From left to right, size of the Hund's coupling $J_{\rm H}$ is enhanced ($J_{\rm H}$ = 0, 0.5, 1.0, 1.1 eV). Note that the high-spin configuration is stabilized at $J_{\rm H}$ = 1.1 eV. 
  }
  \label{figS:T2Eself}
\end{figure}

The inter-site component $\hat{\Sigma}_{ij}$ has a similar form; $\hat{\Sigma}_{ij} \equiv \frac{1}{12} \Sigma^E (\omega) \hat{O}_{ij} $, where
\begin{align}
\hat{O}_{12} &= \left(
\begin{array}{ccc}
-2 & -1 & +1 \\
+1 & +2 & +1 \\
+1 & -1 & -2 
\end{array} \right), 
\hat{O}_{13} = \left(
\begin{array}{ccc}
+2 & +1 & +1 \\
-1 & -2 & +1 \\
-1 & +1 & -2 
\end{array} \right), 
\hat{O}_{23} = \left(
\begin{array}{ccc}
-2 & +1 & -1 \\
+1 & -2 & -1 \\
+1 & +1 & +2 
\end{array} \right), \nonumber \\
\label{eq:offdiag}
\hat{O}_{23} &= \left(
\begin{array}{ccc}
-2 & -1 & -1 \\
-1 & -2 & +1 \\
+1 & -1 & +2 
\end{array} \right), 
\hat{O}_{24} = \left(
\begin{array}{ccc}
+2 & -1 & +1 \\
+1 & -2 & -1 \\
-1 & -1 & -2 
\end{array} \right), 
\hat{O}_{34} = \left(
\begin{array}{ccc}
-2 & +1 & -1 \\
-1 & +2 & +1 \\
-1 & -1 & -2 
\end{array} \right).
\end{align}
Combining (\ref{eq:self1}-\ref{eq:offdiag}), the site-orbital resolved self-energies in the $T^2 \oplus E$ case is as follows. 
\begin{enumerate}[i)]
\item {\it On-site (diagonal blocks), between same orbitals}:     $\left [ \frac{1}{4}\Sigma^{T^2} (\omega)  + \frac{1}{6} \Sigma^E (\omega) \right ]  \hat{I}_{3 \times 3}$,
\item {\it On-site (diagonal blocks), between different orbitals}: $\frac{1}{12} \Sigma^E (\omega) \hat{\Delta}_{ii}$,
\item {\it Inter-site ($i \neq j$ blocks)}: $\frac{1}{4}\Sigma^{T^2} (\omega)  \hat{I}_{3 \times 3} + \frac{1}{12} \Sigma^E(\omega)  \hat{O}_{ij}$.
\end{enumerate}
Here, we note in passing that $\frac{1}{12} \Sigma^E$ is small compared to other terms when $J_{\rm H}$ is not large ($<$ 1 eV), so that terms i) and iii) are dominant contributions, and that the balance between the terms i) and iii) determines whether it is locally (on-site) or non-locally (inter-site) correlated. Plugging (\ref{eq:offdiag}) into the case iii) above yields an explicit expression of the $ij$-block of $\boldsymbol{\Sigma} \left[ T^2 \oplus E \right]$. For example, the block between the site 1 and 2 is as follows,
\begin{align}
\boldsymbol{\Sigma} \left[ T^2 \oplus E \right]_{12} = \left( 
\begin{array}{ccc}
\frac{1}{4}\Sigma^{T^2}{\color{red}\boldsymbol{-}}\frac{1}{6} \Sigma^E & -\frac{1}{12} \Sigma^E &  +\frac{1}{12} \Sigma^E \\
+\frac{1}{12} \Sigma^E & \frac{1}{4}\Sigma^{T^2}{\color{blue}\boldsymbol{+}}\frac{1}{6} \Sigma^E & +\frac{1}{12} \Sigma^E \\
+\frac{1}{12} \Sigma^E & -\frac{1}{12} \Sigma^E & \frac{1}{4}\Sigma^{T^2}{\color{red}\boldsymbol{-}}\frac{1}{6} \Sigma^E 
\end{array} \right), 
\end{align}
where the plus and minus signs in the diagonal components are colored in blue and red to emphasize terms where $\Sigma^{T^2}$ and $\Sigma^E$ are adding up and cancelling out, respectively. 
Among the three diagonal components, the central term ($\frac{1}{4}\Sigma^{T^2}{\color{blue}\boldsymbol{+}}\frac{1}{6}\Sigma^{E}$) is between the $d_{yz}$ orbitals at V site 1 and 2, which are forming a strong $\sigma$-type direct overlap, while the other two $\frac{1}{4}\Sigma^{T^2}{\color{red}\boldsymbol{-}}\frac{1}{6} \Sigma^E$ are contributing to the $\delta$-like weak overlap between the  $d_{xy,xz}$ orbitals. Interestingly, the inclusion of $\Sigma^E$ (and $J_{\rm H}$) affects the inter-site self-energies in an opposite way depending on the orbitals; while the imaginary part of $\frac{1}{4}\Sigma^{T^2}{\color{blue}\boldsymbol{+}}\frac{1}{6}\Sigma^{E}$ is enhanced by the nonzero $\Sigma^{E}$ (because causal self-energies should always have negative imaginary parts), it is canceled out in $\frac{1}{4}\Sigma^{T^2}{\color{red}\boldsymbol{-}}\frac{1}{6}\Sigma^{E}$. This implies that the presence of $\Sigma^{E}$ selectively enhances the singlet moment formation within the stronger $\sigma$-bonding, while reducing inter-site correlations in other bondings. In addition, depending on the sign of the real parts of $\Sigma^{T^2}$ and $\Sigma^{E}$, one can either enhance or suppress the real part of the self-energy.

Fig.~\ref{figS:T2Eself} show the evolution of ${\Sigma}^{T^2,E}(\omega)$ as a function of the Hund's coupling $J_{\rm H}$. Note that the relative signs of the real part of ${\Sigma}^{T^2,E}(\omega)$ tend to be opposite when $J_{\rm H}$ is small, but increasing $J_{\rm H}$ drives them to be the same. Just after the crossover to the high-spin state happens ($J_{\rm H}$ = 1.1 eV), both the ${\rm Re} {\Sigma}^{T^2,E}(\omega)$ show very similar behavior. This is because of the development of the pole in ${\Sigma}^{E}$, signaling the formation of the $E$ local moments, as shown in the lower panels of Fig.~\ref{figS:T2Eself}. As the system goes into the high-spin configuration, both the ${\rm Im} {\Sigma}^{T^2,E}$ should similarly show a well-defined pole, then the shapes of ${\rm Re} {\Sigma}^{T^2,E} (\omega)$ should become similar to each other because of the Kramers-Kronig relation. Hence $\frac{1}{4}\Sigma^{T^2}{\color{red}\boldsymbol{-}}\frac{1}{6} \Sigma^E$ within $\boldsymbol{\Sigma} \left[ T^2 \oplus E \right]_{ij}$ tends to cancel better as $J_{\rm H}$ becomes larger. Since the diagonal parts of the inter-site self energies are most dominant contributions, and we have two $\frac{1}{4}\Sigma^{T^2}{\color{red}\boldsymbol{-}}\frac{1}{6} \Sigma^E$ terms compared to just one $\frac{1}{4}\Sigma^{T^2}{\color{blue}\boldsymbol{+}}\frac{1}{6} \Sigma^E$, the overall self-energy correction to the inter-site hopping terms becomes weaker as the Hund's coupling becomes enhanced. This is consistent with the observation in the main text that increasing $J_{\rm H}$ suppresses the degree of V$_4$ clustering, and that while $U$ enhanced the inter-site correlation via $\Sigma^{T^2}$, $J_{\rm H}$ reduces it by introducing $\Sigma^E$ that cancels $\Sigma^{T^2}$ out. 

\end{document}